\documentclass[twocolumn,showpacs,preprintnumbers,amsmath,amssymb]{revtex4}
\usepackage{graphicx}
\usepackage{dcolumn}
\usepackage{bm}

\begin{document}

\title{\bf Universal Aspects of Curved, Flat  \& Stationary-State Kardar-Parisi-Zhang Statistics}

\author{Timothy Halpin-Healy and Yuexia Lin}
\affiliation{Physics Department, Barnard College, Columbia University, New York NY 10027}
\date{\today}

\begin{abstract}
Motivated by the recent exact solution of the {\it stationary-state} Kardar-Parisi-Zhang (KPZ) statistics by Imamura \& Sasamoto (Phys. Rev. Lett. {\bf 108}, 190603 (2012)), as well as a precursor experimental signature unearthed by Takeuchi (Phys. Rev. Lett. {\bf 110}, 210604 (2013)), we establish here the universality of these phenomena, examining scaling behaviors  of directed polymers in a random medium, the stochastic heat equation with multiplicative noise, and kinetically roughened KPZ growth models. We emphasize the value of cross KPZ-Class universalities, revealing crossover effects of experimental relevance. Finally, we illustrate the great utility of KPZ scaling theory by an optimized numerical analysis of the Ulam problem of random permutations.

\end{abstract}
 


\maketitle

Extremal paths through random energy landscapes, i.e.,  ``directed polymers in random media" (DPRM), have long been a topic of great interest to statistical physicists, condensed matter theorists, and mathematicians alike~\cite{HHZ}.  In two dimensions, the exact limit distributions of the DPRM problem~\cite{SS} are of the celebrated Tracy-Widom (TW) type~\cite{TW}, best known perhaps from random matrix theory~\cite{Mehta,KK}, as well as the famous Ulam problem of random permutations~\cite{Ulam, BB,VK,OR,RSK,BDJ}.  Nevertheless, in the bulk, the statistics of these extremal trajectories remains a challenging, rich, and quite difficult problem~\cite{HH123}.
In all dimensions, however, the constrained free-energy $F({\bf x},t)$ of these directed, extremal paths is dictated by the stochastic partial differential equation~\cite{KPZ} of Kardar, Parisi, \& Zhang (KPZ):
$$\partial_tF=\nu\partial_{\bf x}^2 F +{1\over 2}\lambda(\partial_{\bf x}F)^2 + \surd D \eta,$$
\noindent where $\nu,\hspace{0.5mm}\lambda\hspace{0.5mm}$ and $D$ are system-dependent parameters, the last setting the strength of the {\it additive} stochastic noise $\eta$.  Rigorous mathematical approaches, by contrast, have focussed on the related stochastic heat equation (SHE)
with {\it multiplicative} noise, obtained from KPZ via a Hopf-Cole transformation,   $F$=$\frac{2\nu}{\lambda} {\rm ln}Z$; here, with $\surd\epsilon=\lambda\surd 2D/\nu^3$ and It\^o interpretation, this becomes an elemental, rescaled version~\cite{BC} of the SHE:
$$\partial_t Z=\nabla^2 Z + \frac{\epsilon}{2} Z + \surd\epsilon Z \eta,$$
\noindent
governing the DPRM partition function $Z({\bf x},t)$, a much better behaved quantity, well-regularized in the UV.

Inspired by Imamura \& Sasamoto's exact solution~\cite{IS} of the KPZ equation {\it stationary-state} (SS) statistics, as well as Takeuchi's subsequent re-examination~\cite{KT13} of the tour-de-force KPZ turbulent liquid crystal experiments~\cite{TS}, we focus here on the 1+1 dimensional DPRM/SHE, making manifest its connection to the 
underlying Baik-Rains (BR) F$_0$ limit distribution~\cite{BR} relevant to this KPZ Class.  We examine, too,  DPRM scaling phenomena in the pt-line \& pt-pt configurations, analog of KPZ stochastic growth in {\it flat} \& {\it curved} geometries, governed by Tracy-Widom distributions appropriate to Gaussian orthogonal \& unitary matrix ensembles; i.e., TW-GOE \& GUE, respectively. Our DPRM/SHE results in this regard
provide the final installment of the KPZ triumvirate, complementing prior numerical confirmation of Tracy-Widom universality which had focused initially on kinetic roughening models such as polynuclear-growth (PNG)~\cite{PS},  as well as single-step, or totally asymmetric exclusion processes~\cite{KJ,FF}.  In fact, we go further here in the DPRM/SHE context, making additional suggestions regarding experimental signatures of KPZ Class statistics, examining the interplay of transient \& SS regimes and, finally, in the pt-pt TW-GUE setting, illustrate the potency of KPZ scaling theory~\cite{HS} to inform key aspects of the purely mathematical Ulam problem.  We thus return the favor here, reiterating the very fruitful dialog between KPZ physics and TW GUE mathematics. 

In Figure 1, for starters, we compare to TW-GUE the shifted, rescaled distributions of our {\it curved} KPZ Class models, among them the: i) pt-pt g$_1$ DPRM; i.e., extremal paths in the xy-plane, traveling in the [01]-direction, Gaussian random energies $\varepsilon_i$ with zero mean \& variance $\frac{1}{4}$ on the sites of the square lattice;  side-steps are permitted, but incur a microscopic, elastic energy cost $\gamma$=1. We consider extremal paths that commence at the origin, constrained to terminate t=300 steps later on the axis.  The g$_1$ DPRM trace in Fig 1 represents an average over 10$^8$ realizations of the random energy landscape.  Similarly, ii) pt-pt w/e DPRM- here, the random site energies are exponentially distributed, p($\varepsilon$)=e$^{-\varepsilon}$, and the paths, on average, cut diagonally through the plane in the [11] direction;  at each step, there are two possibilities, either vertically (i.e., in y-direction) or horizontally (x-direction). There is no elastic energy cost in this model, the total energy of the trajectory simply the sum, $F(t)$=$\sum\varepsilon_i$, of the random energies collected traveling from (0,0) to (t,t).  For our w/e DPRM, we averaged over 4x10$^8$ paths of length t=400.  Next, the numerically demanding iii)  constrained SHE-It\^o integrations, with KPZ parameter $\surd\varepsilon$=10; we evolved the system through the forward \& rear light cones of the initial \& final points of the trajectory, averaging over $10^7$ runs, considering paths of 4000 steps, with time increment $\delta$t=0.05. 

\begin{figure}
\includegraphics[angle=0,width=85mm]{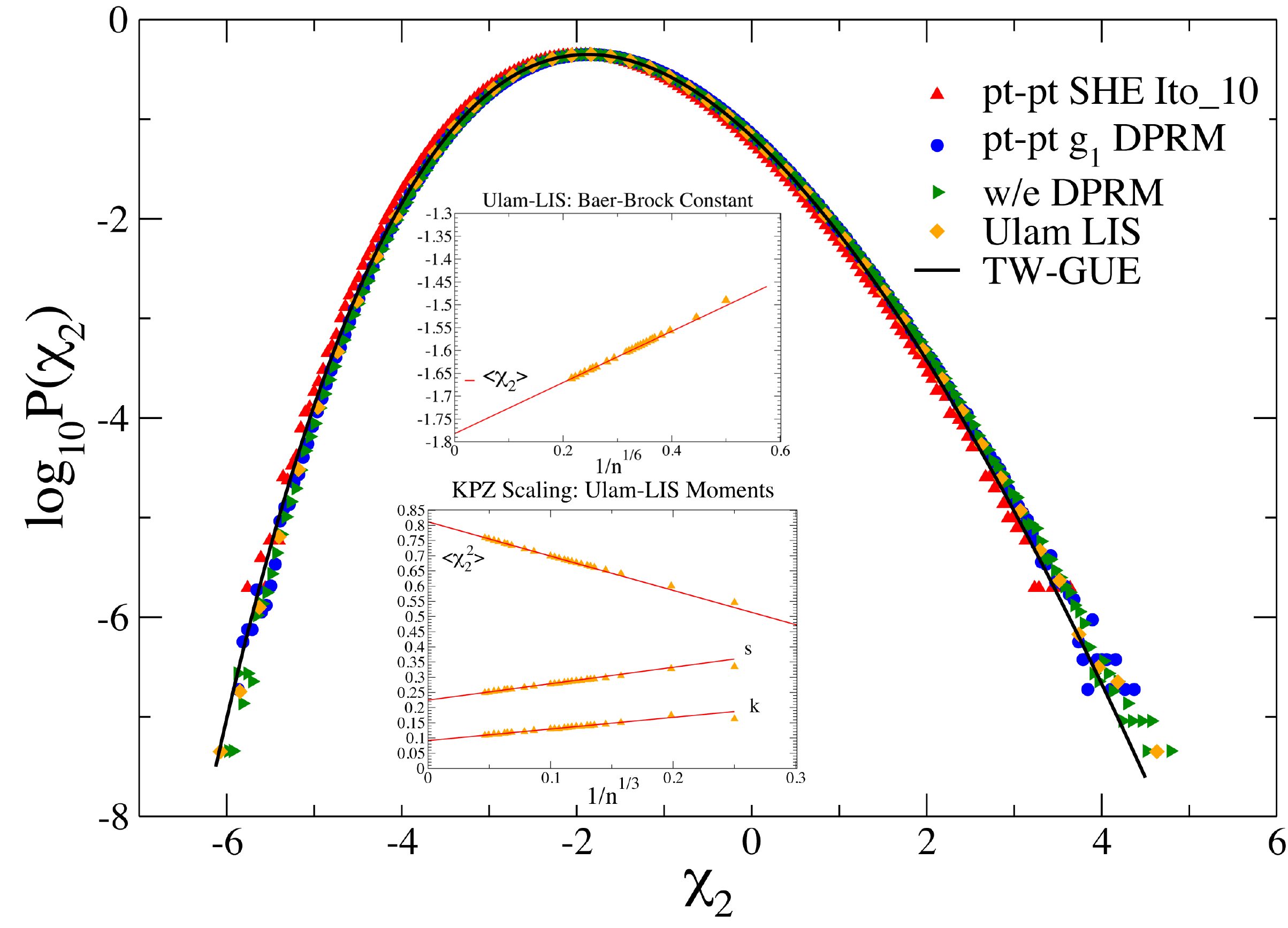}
\caption{\label{fig:Fig1} Universal Limit Distributions: DPRM/SHE, Ulam-LIS, \& TW-GUE. Insets:  Extraction of Ulam-LIS asymptotic moments \& Baer-Brock constant, via KPZ scaling theory.}
\end{figure}

In the case of the pt-pt DPRM models, we have extracted, from a 
first principles Krug-Meakin (KM) finite-size scaling analysis~\cite{KM}, the nonuniversal, {\it system-dependent} KPZ parameters $f_\infty$, $A$=$D/2\nu$, and $\lambda$. Note that the KM toolbox~\cite{note1} yields the asymptotic free energy per step, $f_\infty$, and the {\it product} $A\lambda$, so that $\lambda$, itself, must be fixed by the parabolic DPRM free-energy profile~\cite{note2}.   Interestingly,  in the following, only $f_\infty$ and the combination $\Gamma$=$\frac{1}{2}A^2\lambda$ are needed, which we record in Table I. 
An alternative approach, which highlights the utility of cross KPZ-class studies emphasized in this paper, involves fixing the KPZ scaling parameter $\Gamma$ directly from a fit, e.g.,  to the TW-GOE variance and then {\it using it in the service of} both the TW-GUE limit distribution, our immediate focus, as well as stationary-state Baik-Rains PDF, which we consider later.  This we do, for illustrative purposes, in the case of SHE-It\^o integration.  With $\beta_{KPZ}$=$\frac{1}{3}$ the known~\cite{HHF} DPRM free-energy fluctuation exponent, the central KPZ scaling ansatz reads- $F_{pt-pt}=f_\infty t +(\Gamma t)^\beta\chi_2$ with $\chi_2$ the underlying, order one Tracy-Widom GUE statistical variable.
Thus, the pt-pt DPRM/SHE distributions of Figure 1 have been cast in terms of $\chi_2$, the canonical quantity intrinsic to the {\it curved} KPZ Class in this dimension.  While crafting Figure 1, we have relied upon the model-dependent {\it asymptotic} values for the distribution mean and variance, $\langle\chi_2^2\rangle$=$\lim_{t\rightarrow\infty}\langle\delta F^2\rangle/(\Gamma t)^\beta$, our curve fits for the latter assuming the characteristic  $t^{-2\beta}$ DPRM finite-time correction. Our most accurate values for these limit distribution characteristics are obtained via the Gaussian \& exponential polymers, see Table I; compare, too, 
our model estimates for the distribution skewness $s$=$\langle\delta F^3\rangle_c/ \langle\delta F^2\rangle_c^{3/2}$ and kurtosis $k$=$\langle\delta F^4\rangle_c/ \langle\delta F^2\rangle_c^2.$
The accepted values~\cite{Bourne} for these TW-GUE quantities are: $(\langle\chi_2\rangle,\langle\chi_2^2\rangle,s_2, k_2)$ = $(-1.771087,0.813195,0.224084,0.093448).$  Our pt-pt DPRM-SHE results in  Figure 1, stacked up against the exact, known 1+1 curved KPZ Class TW-GUE distribution, provide an interesting extremal path counterpoint to efforts on radial 2d Eden clusters~\cite{Alves,TSM} and kinetic roughening in droplet geometries~\cite{PS,FF,Alves2}, the latter studies involving PNG, single-step, ballistic deposition (BD), as well as corner growth in a ``restricted-solid-on-solid" (RSOS) model. 

\begin{table*}[!hb]
\hfill{}
\begin{tabular}{l*{7}{c}}
\hline\hline
  KPZ System    & $f_\infty$ & $\Gamma^\beta$ & $\langle\chi_2\rangle$ &  $\langle\chi_2^2\rangle$ & $\vert s_2\vert$ & $k_2$ \\
\hline
g5$_{1}$ DPRM  & -0.498021  & 0.50633 & -1.77097 & 0.8167 & 0.2304 & 0.0924  \\
w/e DPRM  	    &  -2.0003  & 1.9882 & -1.77102 & 0.8116 & 0.2227 & 0.0908  \\
SHE-It\^o   	    & 2.48533 & 0.13345  & -1.7908 & 0.8121 & 0.2263 & 0.0893   \\
\hline\hline
\end{tabular}
\hfill{}
\caption{{\it pt-pt} DPRM/SHE: Model-dependent parameters; estimates of universal TW-GUE quantities.}
\label{tb:KPZp}
\end{table*}

Finally, in homage to the pioneering efforts of Baer \& Brock~\cite{BB}, as well as the extraordinary numerical work of Odlyzko \& Rains~\cite{OR} a generation later,  we have performed, in a post-KPZ/TW context, our own analysis of the Ulam problem, studying the fluctuating statistics of the length, $\ell_n$, of the longest increasing subsequence (LIS) in a permutation of $n$ integers.  
It was established early on, with much heavy-lifting and the securing of mathematical bounds, but then definitively by Vershik \& Kirov~\cite{VK}, that asymptotically  $\langle\ell_n\rangle\rightarrow 2\surd n$.  Indeed, this was well-presented by Baer \& Brock, see Figure 1 and Table 2 of their early paper, where they recorded exact enumerations of the LIS PDF for $n\le36$ using the RSK correspondence, hook formula, \& Young tableau mapping~\cite{RSK}, complementing these results with approximate Monte Carlo data reaching up to $n$=10$^4$. Later, it was conjectured by Odlyzko \& Rains, then firmly established by Kim, that the standard deviation about this mean scaled as $n^{1/6}$; i.e., the Ulam-LIS problem possessed a well-defined limit distribution in the variable $(2\surd n -\ell_n)/n^{1/6}$. Baik, Deift, \& Johansson then proved rigorously that the underlying, asymmetric non-Gaussian distribution was, surprisingly, TW-GUE from random matrix theory~\cite{BDJ}.  In fact, Odlyzko \& Rains provided explicit Monte Carlo evidence in this regard, matching their semilog LIS PDF, with 10$^5$ permutations of length $n$=10$^6$, against a precise Painlev\'e II rendering of Tracy-Widom GUE.  Interestingly, the Odlyzko-Rains data show a small, but clear, finite-time offset (they quote $\langle\chi_2\rangle$$\approx$--1.720) in their LIS PDF, which only stubbornly disappears as the authors heroically head toward asymptopia, pushing their simulations to $n$=10$^8$, where $\langle\chi_2\rangle$$\approx$--1.758,  \& beyond ($n$=10$^{10}$), nearly closing the gap between themselves and the known TW-GUE universal mean: $\langle\chi_2\rangle$=--1.7711.  Since $\beta_{LIS}$=$\frac{1}{6}$, random permutations of $n$=10$^{10}$ digits are equivalent, roughly, to generating pt-pt DPRM with t=10$^5$, quite a demanding task,  indeed.  Here, we take a different attitude, considering rather short permutation strings, with $n$=10$^2$-10$^4$, essentially the regime where Baer \& Brock concentrated their numerical enterprise, but invoke KPZ scaling wisdom to, nevertheless, pin down very precise values for the moments; see Fig. 1, where our Ulam-LIS PDF follows from 10$^8$ random permutations of length $n$=10$^4$, as well as the figure insets, which document the KPZ scaling of the truncated LIS cumulants, allowing us to make fine asymptotic estimates: (-1.7715,0.8135,0.2245,0.09102), close to known TW-GUE values. For the Ulam-LIS variance, skewness \& kurtosis, lower inset, we've invoked an $n^{-2\beta}$ finite-``time" correction, suggested by DPRM lore. The upper inset, which examines KPZ scaling of the universal TW-GUE mean, amends the Odlyzko-Rains ansatz to include an additive constant term, resulting in an $n^{-\beta}$ finite-time correction to $\langle\chi_2\rangle$, plainly visible in the data. In our Ulam-LIS work, we've come to refer to this quantity as Baer-Brock's constant- here estimated to be $\approx$0.51$\pm$0.01, given by the slope of the line in our plot of  $\langle\chi_2\rangle$ versus $n^{-1/6}.$ The existence of such a {\it model-dependent additive constant} is well-known to KPZ practitioners~\cite{SSNP,FF,Alves2,Kat}, but has not, to our knowledge, been discussed for Ulam-LIS. Given the somewhat priviliged role of the random permutation problem amidst the menagerie of systems obeying TW-GUE statistics- it is, by the hook formula, uniquely accessible to exact enumeration for finite $n$- we attach somewhat greater significance to this particular additive constant.
 
\begin{figure}
\includegraphics[angle=0,width=85mm]{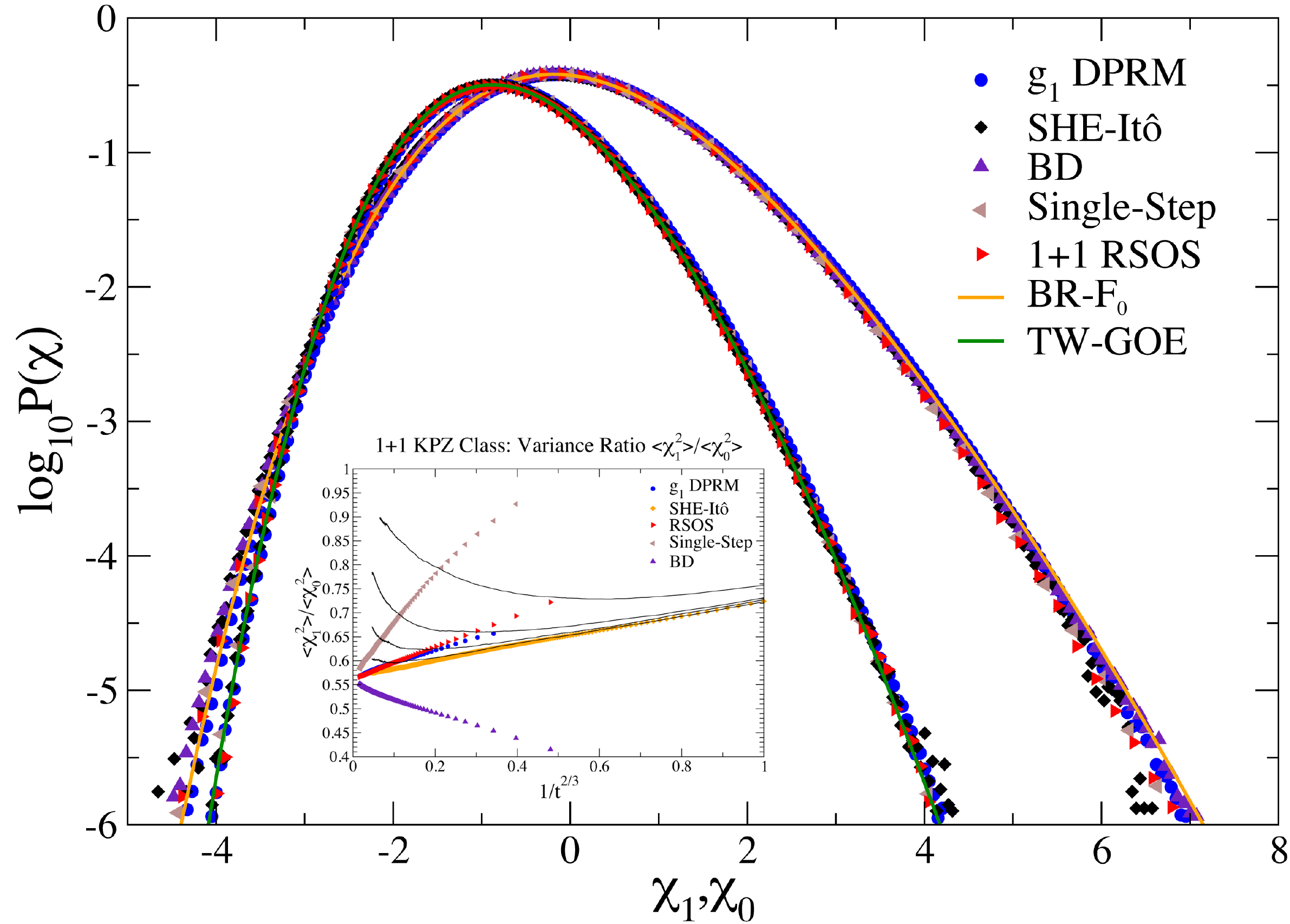}
\caption{\label{fig:Fig2}  1+1 {\it transient-regime} \& {\it stationary-state} KPZ/DPRM Class: Comparison to TW-GOE \& BR-F$_0$ limit distributions.
Inset: Universal {\it variance-ratio} as precursor signature of KPZ SS statistics. Black traces document crossover scaling to the stationary-state,
accessible via experiment. }
\end{figure}   
 
In Figure 2, for our g$_1$ DPRM, single-step growth, and SHE-It\^o Eulerian integration, we have paired up the distinct PDFs associated with the {\it transient-regime} \& {\it stationary-state} statistics for the flat KPZ geometry, making relevant comparisons to exact Tracy-Widom GOE \& Baik-Rains F$_0$ distributions, respectively.  For the {\it transient} dynamics, our DPRM/SHE work harks back to early, pre-TW efforts~\cite{KBM,HH91}, and dovetails with recent growth model studies~\cite{Olive}, in which several KPZ kinetic roughening PDFs, among them BD \& RSOS, are fit to TW-GOE.  As an indication of our findings here, we note specifically that the g$_1$ DPRM yields a variance, 0.63703, quite close to the TW-GOE result $\langle\chi_1^2\rangle$=0.63805, along with a skewness, 0.2976, very much in line with the value, $s_1$=0.2935, characteristic of that distribution.   More importantly, with regard to 1+1 KPZ Class {\it stationary-state statistics,} and comparison to  the Baik-Rains limit distribution, our DPRM/SHE results, well-matched by single-step, BD, \& RSOS findings,  convincingly establish universality of the seminal PNG studies of Pr\"ahofer \& Spohn~\cite{PS}- note the strong data collapse  of our models upon the BR-F$_0$ trace, a highly nontrivial numerical matter.  For the g$_1$ DPRM, we have evolved the system to a late time t$_o$=8x10$^4$, then study dynamic temporal correlations in the path {\it free-energy,} $\Delta F(t)$=$F(x,t_o+t)-F(x,t_o)$, during the subsequent time interval $t$=500, averaging over 8x10$^4$ runs in a system of size $L$=10$^4$. At $t$=500, we note the approximate values: (1.1177,0.3432,0.252) which, extrapolated to infinite $t$, yield (1.1466,0.3530,0.278), quite close to the known~\cite{PS} BR-F$_0$ variance, skewness, \& kurtosis: ($\langle\chi_0^2\rangle$,$s_0$,$k_0$)=(1.15039,0.35941,0.28916).  For RSOS growth in the stationary-state, $t_o$=40k \& $t$=240, which gives an asymptotic estimate $\langle\chi_0^2\rangle\approx$1.1331 for the Baik-Rains constant~\cite{chi2}, while ballistic deposition, with $t_o$=50k \& $t$=500, produces 1.1382 for this quantity.  We emphasize that the plotted distributions of Figure 2 are not the raw data, but have been rescaled, each in turn, to represent the {\it system-dependent} asymptotic variances, carefully extracted via KPZ scaling theory.  

Included as Fig 2 inset, we consider the time dependence of the {\it ratio of variances} in KPZ transient \& stationary-state regimes.  
We are motivated here by Takeuchi's recent study~\cite{KT13} of the PNG model, as well his attempts to tease from the 1+1 KPZ Class turbulent liquid-crystal data~\cite{TS}, some sign of the SS statistics.  Although the BR-F$_0$ limit distribution itself remained well-beyond the reach of these experiments, Takeuchi discovered an impressive precursor signature of the KPZ stationary-state, evident as a {\it skewness minimum,} which we confirm shortly with our own models, establishing its universal KPZ aspect.  Given the intrinsic importance of the stationary-state dynamics~\cite{IS,PS}, we propose here an alternative signature, relying upon a more statistically tame quantity which, asymptotically, is fixed universally-
$$\lim_{t\rightarrow\infty}\langle\delta F^2(t)\rangle_{tran}/\langle\Delta F^2(t)\rangle_{SS}=\frac {\langle\chi_1^2\rangle} {\langle\chi_0^2\rangle}=0.55464$$
by the ratio of the TW-GOE variance \& Baik-Rains constant.   Since the {\it variance-ratio,} like the skewness $s$, requires no knowledge of $\Gamma$, a quick examination of this quantity in experiments may provide a glimpse of a key universal property of the KPZ stationary-state.  Within Figure 2 inset, we document (solid symbols) the scaling of this interesting ratio for g$_1$ DPRM, SHE-It\^o, and three distinct KPZ growth models. All 5 systems converge convincingly- even the contrarian BD, well known for its curmudgeonly behavior~\cite{DVD}.   For the SHE-It\^o case, $t_o$=10k \& $t$=100 as above, per our rendering of the BR distribution.  However, we also include within the inset, data sets associated with lesser values of $t_o$ \& hundredfold fewer runs, to study {\it crossover effects,} indicating how this scaling phenomena might appear in a more constrained experimental context. The black traces, top down, correspond to $t_o$=3, 20, 100, 500, though all possess $t$=100, with statistical averaging now done~\cite{vrnote} over a 10$^6$, rather than 10$^8$ points.  The goal here is to provide an extra tack upon the 1+1 KPZ Class {\it stationary-state} statistics, and to further complement the spectacular experimental results already in hand for the transient {\it flat } TW-GOE and {\it curved} TW-GUE KPZ Classes. 
 
\begin{figure}
\includegraphics[angle=0,width=85mm]{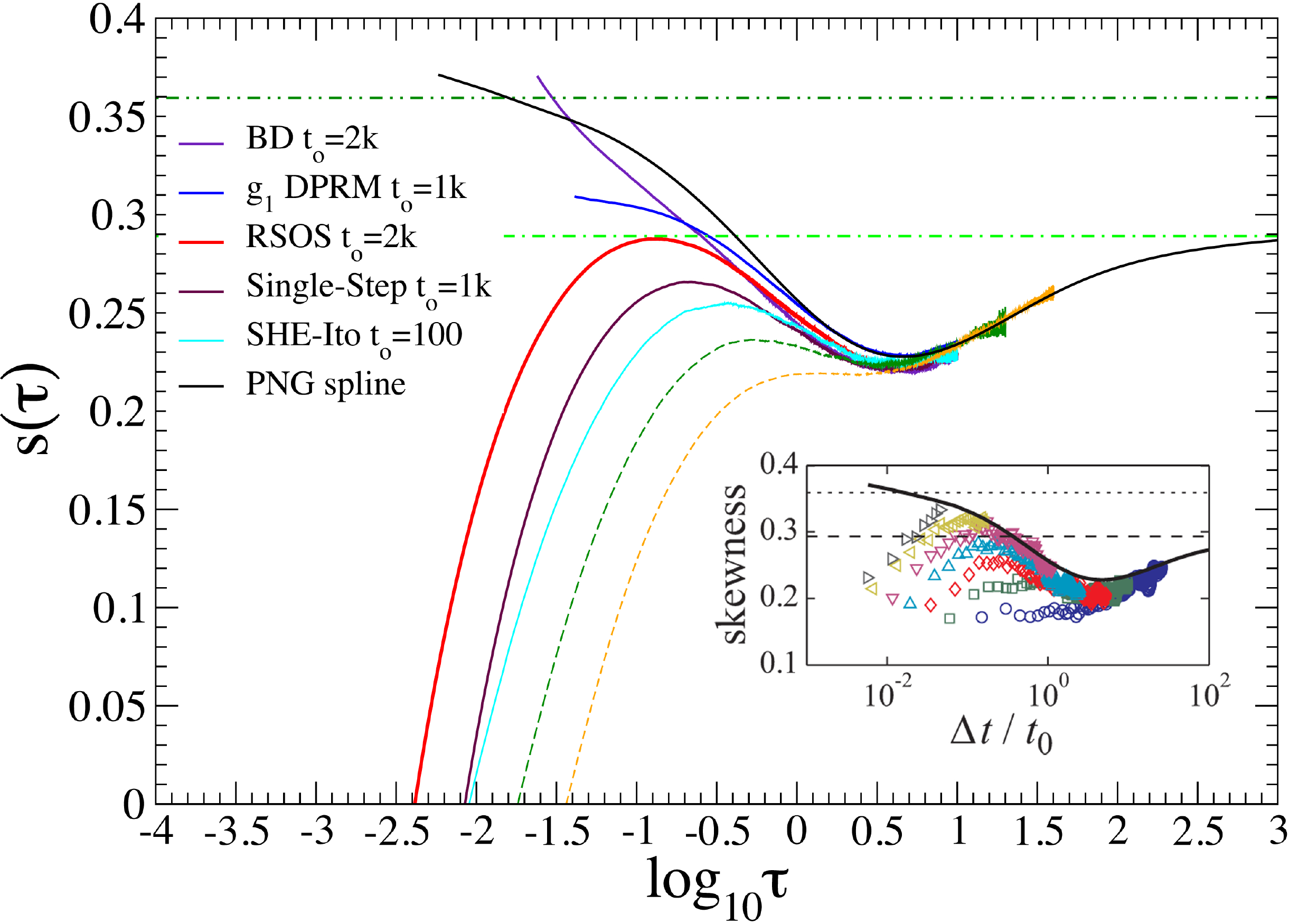}
\caption{\label{fig:Fig3} Takeuchi {\it skewness minimum}, evident in our DPRM, SHE-It\^o, BD, RSOS \& Single-Step kinetic roughening models.
Upper \& lower dot-dashed lines correspond to asymptotic skewness, $s_0$=0.35941 \& $s_1$=0.2935, of the Baik-Rains F$_0$ \& Tracy-Widom GOE limit distributions, respectively.
Inset: Data sets, KPZ turbulent liquid crystal experiments- Ref.~\cite{KT13}. The emergence of the minimum for our SHE-It\^o integration,
revealed as $t_o$ grows from 25, 50, to 100, is manifest within the experiment proper, where the symbols indicate distinct times, smallest to largest, $t_o$=2,6,10,18,30,54,60, right to left.} 
\end{figure}

Lastly, with Figure 3, we establish {\it universality} of Takeuchi's skewness minimum~\cite{KT13}, heightening its utility as an experimental precursor signature of the 1+1 KPZ Class stationary-state statistics. For our Gaussian DPRM, SHE-It\^o integration, and KPZ growth models, we have plotted up the fluctuation PDF skewness $s$ as a function of the dimensionless scaling parameter $\tau$=$\Delta t/t_o$ where, in each case, the numerics are evolved to a time $t_o$, the skewness then tracked through a subsequent time interval $\Delta t$.  To consider the consequences of an {\it insufficiently mature} kinetically-roughened state, we include SHE-It\^o results for $t_o$=25, 50, \& 100 respectively, shown in the ascending, dashed curves, solid in the last instance.  It is interesting to observe how, with increasing $t_o$, 
the traces develop to reveal a nearly full-fledged Takeuchi minimum; i.e., for $t_o$=25, the skewness min is entirely absent, though the curve possesses a suggestive inflection point.  At $t_o$=50, the minimum has appeared, but only gains a fuller expression at $t_o$=100.   Note, however, that once the skewness minimum has emerged, there is very little movement in its precise location- this is especially true of the $s_{TM}$ ordinate, slightly less so for its abscissa.  The evolution, with increasing $t_o$, of the universal KPZ Class skewness minimum is tied primarily to its {\it curvature} and, in the 1+1 dimensional case, to the steepening slope for $\tau\lesssim\tau_{TM}$, as $t_o\rightarrow\infty$.  We mention, in passing, that our SHE-It\^o, single-step, and RSOS simulations share a common {\it nonuniversal behavior} at very early $\Delta t$, which gives rise to a {\it model-dependent} maximum and subsequent ($\tau\ll1$) monotonic decrease in $s(\tau)$, evident in the figure.  Most interestingly, this peculiar feature is actually manifest within Takeuchi's~\cite{KT13} liquid-crystal KPZ experimental data set proper(!), shown as an inset within Figure 3 and contrasts, intriguingly, with his own PNG simulation results, which indicate a weakly divergent, {\it increasing} $s$ for the smallest $\tau$ values~\cite{final}.  These system-dependent details are small, but curious matters, slightly off-stage, but  nevertheless quite apparent to those involved in the nuts \& bolts of simulation.  From an experimental point of view, however, the Takeuchi minimum itself is the essential focus and sits center-stage, providing a {\it crucial, tell-tale precursor signature of 1+1 KPZ Class stationary-state statistics.}  Hence, see Figure 3, our g$_1$ DPRM results, for which we have devoted the greatest numerical investment, making a 1/4 million runs in a system size L=10$^4$, resulting in a data set of 2.5 billion samples.  Included, too, in the figure itself, as well as the inset, is Takeuchi's skewness spline (black traces), carefully crafted via his PNG work.  The agreement between Takeuchi's PNG spline and our mix of KPZ data sets, and that of the g$_1$ DPRM in particular, is quite suggestive, indeed.  From our simulations, we locate the Takeuchi minimum at  $s_{TM}$=0.225$\pm$0.005, for $\tau$=$\frac{\Delta t}{t_o}$=4.5$\pm$0.4.  We note, that the nonuniversal behavior at small $\tau\ll1$ evident in Takeuchi's PNG simulation, giving rise to a climbing, positive $s$ is, in fact, shared by our g$_1$ DPRM and BD models.  Of course, it is only in the double limit, $t_o\rightarrow\infty$, then $\Delta t\rightarrow\infty$, with $\tau=\Delta t/t_o\rightarrow0$ that an interpolating skewness spline, originating at the TW-GOE value, $s_1$=0.2935, for large $\tau\gg 1$, would descend through the Takeuchi minimum at the sweetspot near $\tau\gtrsim 1$, then exhibit the correct asymptotic approach to the stationary-state Baik-Rains value, $s_0$=0.35941, for vanishing small $\tau\ll 1$. To manifest the latter would be quite difficult, even numerically.  Finally, we mention that the  {\it kurtosis minimum} we find for the  g$_1$ DPRM, at $k_{TM}$=0.117, is quite near Takeuchi's PNG value for this more demanding, higher cumulant ratio.  Given the tougher statistics, as well as the smaller difference here between the TW-GOE kurtosis, $k_1$=0.1652, and our measured k$_{TM}$, eliciting this delicate feature experimentally will remain, no doubt, a most challenging task.  In the interim, 
we await a clever experimental implementation of the KPZ {\it stationary-state} initial condition, which might allow {\it direct} access to the Baik-Rains F$_0$.  Given the universality established via Figure 2, with our DPRM/SHE, single-step, BD, and RSOS results, this limit distribution represents, most assuredly, the relevant fixed point PDF.
\section{\bf Acknowledgements}
\vspace{-2mm} 
Many thanks to K. A. Takeuchi for numerous fruitful exchanges regarding our work and for providing his PNG skewness splines \& 1+1 KPZ Class experimental data. We're very grateful, as well, to M. Pr\"ahofer for making available the TW-GUE \& GOE traces, and to T. Imamura for kindly sharing his numerical rendering of the Baik-Rains F$_0$ limit distribution.


\begin{thebibliography}{10}
\bibitem{note2}  In the case of 1+1 KPZ stochastic growth models, the parameter $A$ can be determined by the {\it steady-state} width of the interface, which scales with the finite system size $L$ via the relation $w$=$\surd AL/12.$  Alternatively, the KPZ nonlinearity $\lambda$ is fixed by the tilt-dependent growth velocity: $\delta v$=$\frac{1}{2}\lambda(\delta h/\delta x)^2$; these matters are amply discussed by Krug, Meakin, \& Halpin-Healy~\cite{KM}.
\bibitem{vrnote} We are imagining, here, an experimental system size $L$=10$^3$ pixels, with roughly 10$^3$ runs being made,
yielding 10$^6$ data for each time slice.
\end{thebibliography}
\end{document}